# Slow periodic oscillations in time domain dynamics of NO$_2$

Michaël SANREY and Marc JOYEUX[a]

*Laboratoire de Spectrométrie Physique (CNRS UMR5588),*

*Université Joseph Fourier, BP 87, 38402 St Martin d'Hères, France*

**Abstract :** We investigated the time domain nonadiabatic dynamics of NO$_2$ on the coupled X$^2$A$_1$ and A$^2$B$_2$ electronic states by launching wave packets on the excited electronic state and focused on the evolution at long times ($t > 200$ fs), which has received little attention up to now. We showed that the initial fast spreading of the wave packets is followed at all energies by slow periodic intramolecular vibronic energy redistribution (IVER) with periods in the range 0.3 to several tens of ps. These energy transfers lead to oscillations with the same periods in the population of each electronic state. Propagation of wave packets indicates that IVER frequencies also dominate the fluctuations of the squared modulus of the autocorrelation function, $|A(t)|^2$, at energies not too high above the bottom of the conical intersection, but that this is no longer the case at higher energies. For example, for initial wave packets prepared by almost vertical excitation of the vibrational ground state of the ground electronic surface, the oscillations of $|A(t)|^2$ essentially reflect the detuning from 1:2 resonance between the frequency of the bend and that of the symmetric stretch in the excited electronic state. These theoretical results were used to discuss the possible origin of the low frequency oscillations which were recently observed in time domain experimental spectra of NO$_2$.

(a) email : Marc.JOYEUX@ujf-grenoble.fr

## I - Introduction

NO$_2$ allies the simplicity of a triatomic molecule - and the associated amenability to sophisticated calculations and high resolution experiments - to a rich and complex behaviour that originates mainly from the $X^2A_1$-$A^2B_2$ conical intersection between the two lowest electronic states of the molecule. It has consequently served as a benchmark for experimental and theoretical studies dealing with vibronic coupling and non-Born-Oppenheimer dynamics, both in the frequency (see for example Refs. [1-12]) and time (see for example Refs. [13-20]) domains. While most essential questions pertaining to the frequency domain seem to be now well understood, several fundamental points dealing with time domain properties still deserve detailed investigation. For example, little is known concerning the properties of the long time evolution, although the short time ($t < 200$ fs range) behaviour has been investigated by several authors and is now well characterized [17-20]. The need for additional work in this direction was clearly pointed out by Mahapatra *et al* in their study of nonadiabatic wave packet dynamics on realistic ab initio potential energy surfaces [17]. They indeed noticed that the initial fast decay of the $A^2B_2$ diabatic population is followed by "wiggles" that "are not a numerical artefact" since they are "reproduced by varying the grid size". After mentioning that "similar effect has also been found previously in other weakly coupled systems" [15], the authors conclude that "a precise explanation for the time scale of the oscillations cannot be given at present". Quite interestingly, experimentalists who use two colours pump-probe laser set-ups with photoion and/or photoelectron detection to investigate the time-domain dynamics of NO$_2$ close to its first dissociation limit [13,14] come up against the same kind of question : their most intriguing result is indeed the observation of a slow NO$^+$ transient signal with a particularly pronounced oscillatory component of period 600-800 fs, that is, an equivalent energy spacing of about 40-55 cm$^{-1}$. A first study concluded that these oscillations "may measure the average energy level spacing between the resonant levels of the $A^2B_2/X^2A_1$ states close to their conical intersection, but alternative explanations cannot be ruled out" [13]. In contrast, a second study tentatively assigned them to "wave packet motion along a very soft coordinate with an underlying harmonic level spacing of ~40 cm$^{-1}$" and suggested that this motion might be an "essentially free rotation of an oxygen atom around a core NO molecule at large O-NO distances" [14].

It is clear from what precedes that the dynamics of $NO_2$ at long times is still poorly understood. It is precisely the purpose of this paper to fill part of this blank on the basis of the detailed analysis of the dynamics of wave packets propagated on the coupled surfaces. In particular, it will be shown how the wiggles reported in theoretical calculations of the time evolution of the population of the excited electronic state and the experimentally observed periodic oscillations of the $NO^+$ signal may or may not be connected to the nonadiabatic coupling between the two surfaces.

The remainder of this article is organized as follows. Section II contains a description of the model Hamiltonian we used in this work. The time evolution of the excited state population and of the autocorrelation function are analyzed in Section III for wave packets launched at several energies on the excited electronic state. Finally, we discuss in Sect. IV the connection between the present calculations and experimental results [13,14].

**II - The effective model**

The model used in this work is derived from that of Ref. [12], whose parameters were adjusted against the experimentally determined energies of the first 307 vibronic states. In order to investigate the dynamics at much higher energies without having to worry about the behaviour of high-order polynomials at large displacements, most of the anharmonicities of the model were however neglected. More explicitly, the model used here is of the form

$$\mathbf{H} = \begin{pmatrix} H_e & H_c \\ H_c & H_g \end{pmatrix},  \quad \text{(II-1)}$$

where $H_g$ and $H_e$ denote the Hamiltonians for the ground and excited diabatic electronic states, while $H_c$ is the coupling term. $H_g$ is assumed to be harmonic, with matrix elements

$$\langle v_1, v_2, v_3 | H_g | v_1, v_2, v_3 \rangle = \sum_{i=1}^{3} \omega_i (v_i + \frac{1}{2}) \quad \text{(II-2)}$$

in the direct product harmonic basis $|v_1, v_2, v_3\rangle$ of the ground electronic state. The matrix elements of $H_e$ in the direct product harmonic basis $|v'_1, v'_2, v'_3\rangle$ of the excited electronic state have a slightly more complex form

$$\langle v'_1, v'_2, v'_3 | H_e | v'_1, v'_2, v'_3 \rangle = E'_0 + \sum_{i=1}^{3} \omega'_i (v'_i + \frac{1}{2}) + x'_{22}(v'_2 + \frac{1}{2})^2 . \quad \text{(II-3)}$$

In Eqs. (II-2) and (II-3), indexes 1 to 3 refer to the symmetric stretch, the bend and the antisymmetric stretch, respectively. $E'_0$ is the energy gap between the minima of the two electronic surfaces. We will come back later to the role of the $x'_{22}$ anharmonicity in the present study. The coupling $H_c$ is the only first order term authorized by symmetry, that is

$$H_c = \lambda q_3 , \quad \text{(II-4)}$$

where $q_3$ is the antisymmetric stretch dimensionless position coordinate for the ground electronic state. Finally, the sets of dimensionless normal coordinates $\mathbf{q} = (q_1, q_2, q_3)$ and $\mathbf{q'} = (q'_1, q'_2, q'_3)$ in the ground and excited electronic states are connected through

$$\mathbf{q'} = \mathbf{A} \ \mathbf{q} + \mathbf{B} = \begin{pmatrix} 0.899 & -0.532 & 0 \\ 0.301 & 0.906 & 0 \\ 0 & 0 & 0.693 \end{pmatrix} \mathbf{q} + \begin{pmatrix} 1.100 \\ -5.730 \\ 0 \end{pmatrix}, \quad \text{(II-5)}$$

while conjugate momenta satisfy

$$\mathbf{p}' = {}^t\mathbf{A}^{-1}\mathbf{p} .\qquad (\text{II-6})$$

Numerical values of the spectroscopic constants in Eqs. (II-2)-(II-4) are those of Ref. [12], that is $\omega_1 = 1357.4090$ cm$^{-1}$, $\omega_2 = 756.8245$ cm$^{-1}$, $\omega_3 = 1670.3878$ cm$^{-1}$, $E_0' = 10209.0874$ cm$^{-1}$, $\omega_1' = 1315.0927$ cm$^{-1}$, $\omega_2' = 766.2419$ cm$^{-1}$, $\omega_3' = 753.1696$ cm$^{-1}$ and $\lambda = 331.7647$ cm$^{-1}$. In addition, we assumed that $x_{22}' = -3.0$ cm$^{-1}$. The eigenfunctions

$$\Psi_j = \begin{pmatrix} \Psi_j^e \\ \Psi_j^g \end{pmatrix} \qquad (\text{II-7})$$

and energies

$$E_j = \langle \Psi_j | \mathbf{H} | \Psi_j \rangle . \qquad (\text{II-8})$$

of the first 3800 eigenstates of the coupled Hamiltonian were computed according to the method described in Ref. [12].

**III - Wave packet dynamics**

We launched minimum uncertainty wave packets from several positions on the excited electronic state. At time $t = 0$, the component of the wave packets on the ground electronic state, $\Phi^g(0)$, is thus zero, while the component on the excited electronic state, $\Phi^e(0)$, is of the form

$$\Phi^e(0) = \prod_{k=1}^{3} \pi^{-1/4} \exp\left\{ i\,\bar{p}_k' q_k' - \frac{1}{2}(q_k' - \bar{q}_k')^2 \right\} \qquad (\text{III-1})$$

where the $\bar{q}_k'$ and $\bar{p}_k'$ ($k=1,2,3$) denote initial positions and momenta of the wave packet. If $c_j$ is the projection of $\Phi^e(t=0)$ on the excited state component of the $j$-th eigenstate of the coupled Hamiltonian,

$$\Phi^e(0) = \sum_j c_j \Psi_j^e \ , \tag{III-2}$$

then the wave packet at time $t$ may be written in the form

$$\Phi(t) = \begin{pmatrix} \Phi^e(t) \\ \Phi^g(t) \end{pmatrix} = \sum_j c_j \exp(-iE_j t)\Psi_j = \begin{pmatrix} \sum_j c_j \exp(-iE_j t)\Psi_j^e \\ \sum_j c_j \exp(-iE_j t)\Psi_j^g \end{pmatrix}. \tag{III-3}$$

For practical purposes, the population in the excited electronic state,

$$P^e(t) = \langle \Phi^e(t) | \Phi^e(t) \rangle \ , \tag{III-4}$$

is best rewritten in the form

$$P^e(t) = \sum_j |c_j|^2 |\Psi_j^e|^2 + \sum_{j<k} (c_j^* c_k + c_j c_k^*)\langle \Psi_j^e | \Psi_k^e \rangle \cos((E_j - E_k)t) \ . \tag{III-5}$$

Fig. 1 displays the time evolution of $P^e$ for two wave packets which at time $t=0$ are of minimum uncertainty and centred around $p_1 = p_2 = p_3 = q_1 = q_3 = 0$ and $q_2 = 3.0$ (dashed line) or $q_2 = 0.5$ (solid line) on the upper $A^2B_2$ electronic state (use Eqs. (II-5) and (II-6) to get the initial conditions in terms of the normal coordinates of the excited electronic states, *i.e.* the $\bar{q}'_k$ and $\bar{p}'_k$ that appear in Eq. (III-1)). Potential energy at the centre of the wave packets at time $t=0$ is approximately 13800 cm$^{-1}$ for $q_2 = 3.0$ and 20800 cm$^{-1}$ for $q_2 = 0.5$. This means that the centre of the wave packet launched at $q_2 = 3.0$ is located only 3600 cm$^{-1}$ above the minimum of the excited electronic state while that of the wave packet launched at $q_2 = 0.5$ is located 10600 cm$^{-1}$ above this minimum. The time evolution of $P^e$ is seen to consist of two phases. The first phase, which lasts about 200 fs at these energies, corresponds to the initial spreading of the wave packet, which progressively occupies the whole accessible phase space in both electronic states. As described in detail in Ref. [20], this phase is associated with a strong and fast decrease of $P^e$, which occurs as a series of steps. Once this phase is over, the population in each electronic state fluctuates around an average value, in excellent agreement with the results of Mahapatra *et al* obtained with *ab initio* potential energy surfaces [17]. Note that classical models reproduce perfectly well the initial decrease of $P^e$ but display no fluctuations during the second phase of the time evolution [20].

As noticed by Mahapatra *et al* [17], population fluctuations during the second phase look like noise, but they are not noise. This is not surprising, since Eq. (III-5) indicates that the oscillatory part of $P^e$ is the sum of periodic contributions whose angular frequencies $|E_j - E_k|$ are differences between the energies of pairs of eigenstates, while amplitudes are equal to $(c_j^* c_k + c_j c_k^*) \langle \Psi_j^e | \Psi_k^e \rangle$. The squared modulus of the Fourier transform of $P^e(t)$ for the wave packets launched at $q_2 = 3.0$ and $q_2 = 0.5$ is shown in the top plots of Figs. 2 and 3, respectively. The spectrum of the wave packet with lowest energy essentially consists of a series of six strong lines in the 0-120 cm$^{-1}$ frequency range. These lines are labelled A to F in the top plot of Fig. 2. The lines in the spectrum of the wave packet launched at $q_2 = 0.5$ are significantly weaker and appear at comparatively lower frequencies. They are labelled W to Z in the top plot of Fig. 3. The pair of eigenstates at the origin of each of these lines is indicated in the last column of Table I, while the largest terms of their decomposition on the harmonic bases of the ground and excited electronic states are provided in Table II. Note that in this latter table $(v_1, v_2, v_3)$ stands for the vector $|v_1, v_2, v_3\rangle$ of the harmonic oscillator basis of the ground electronic state and $[v_1', v_2', v_3']$ for the vector $|v_1', v_2', v_3'\rangle$ of the harmonic oscillator basis of the excited electronic state. When the two electronic states are decoupled, *i.e.* when $\lambda = 0$, all $\langle \Psi_j^e | \Psi_k^e \rangle$ with $j \neq k$ are zero and $P^e(t)$ is of course constant. Moreover, states which do not contribute to the initial wave packet have projections $c_j \approx 0$, so that they cannot give rise to any line. Therefore, pairs of states which significantly contribute to the fluctuations of $P^e(t)$ necessarily (i) result from the vibronic coupling between at least two vectors of the harmonic bases, and (ii) are significantly populated at time $t=0$. Tables I and II accordingly show that lines A to F borrow their intensity from the vibronic coupling between [0,5,0] and (2,12,1), [0,6,0] and (2,13,1), [0,4,0] and (2,11,1), [0,2,0] and (2,9,1), [0,2,0] and (1,11,1), and [0,4,0] and (1,13,1), respectively, while lines W to Z result from the vibronic coupling between [1,16,0] and (12,6,1), [0,10,0] and (5,7,3), [0,16,0] and (1,24,1), and [0,17,0] and (2,23,1), respectively. This is easily understood from the considerations above and from the fact that excitation at time $t=0$ is almost exclusively localized in the bend degree of freedom, with $v'_2$ centred around 4 for the wave packet with lowest energy and around 14 for the wave packet with highest energy. To summarize, fluctuations of $P^e(t)$ are thus a sensitive fingerprint of the vibronic coupling and reflect the IVER rates of the model.

Let us now consider the time evolution of the squared modulus of the autocorrelation function

$$|A(t)|^2 = |\langle \Phi(t) | \Phi(0) \rangle|^2 , \tag{III-6}$$

which can be written in the form

$$|A(t)|^2 = \sum_j |c_j|^4 + 2 \sum_{j<k} |c_j|^2 |c_k|^2 \cos((E_j - E_k)t) , \tag{III-7}$$

Fig. 4 displays the time evolution of $|A(t)|^2$ for the same wave packets that were used to plot Figs. 1-3. After the initial fast decrease, the evolution of $|A(t)|^2$ consists of fast oscillations superposed on top of slower ones. The slow oscillations become clearer when the signal is averaged with a gaussian window of 90 fs width (FWHM), as can be checked in Fig. 4 where the averaged signals appear as white traces. The squared modulus of the Fourier transform of $|A(t)|^2$ in the range 0-2000 cm$^{-1}$ is shown in Fig. 5 for the two wave packets, while the bottom plots of Figs. 2 and 3 display zooms of the Fourier transforms in the range 0-125 cm$^{-1}$. According to Eq. (III-7), periodic oscillations of $|A(t)|^2$ have angular frequencies that are differences between the energies of pairs of eigenstates, like for $P^e(t)$. The amplitudes of the oscillations are however very different from those of $P^e(t)$, in the sense that it is sufficient that eigenstates $j$ and $k$ be significantly excited at time $t=0$ for the oscillations at $|E_j - E_k|$ to contribute significantly to the evolution of $|A(t)|^2$. In contrast with Eq. (III-5), the scalar product $\langle \Psi_j^e | \Psi_k^e \rangle$ does not appear in the expression of $|A(t)|^2$, so that it is not necessary that states $j$ and $k$ be resonantly coupled. For example, the frequency of the strongest line that appears in the Fourier transform of $|A(t)|^2$ in the range 0-2000 cm$^{-1}$ for the wave packet launched at $q_2 = 3.0$ (line h at 754.28 cm$^{-1}$ in the top plot of Fig. 5) corresponds to the energy difference between eigenstates #272 and #326 (see Table I), which can be described as [0,3,0] and [0,4,0], respectively, and are NOT resonantly coupled (see Table II). Similarly, the frequency of the very strong line at 1478.33 cm$^{-1}$ (labelled k) corresponds to the energy difference between eigenstates #326 and #386 (see Table I), where state #386 can be described as [0,5,0]. Examination of Table II again indicates that these two states are NOT resonantly coupled. Thus, the two bands that appear in the Fourier transforms of $|A(t)|^2$ in the range 0-2000 cm$^{-1}$ have frequencies that are approximately $\omega_2'$ and $2\omega_2'$. In addition to the harmonics of $\omega_2'$, the Fourier transform of $|A(t)|^2$ also displays several lines close to the origin, as can be seen in the bottom plots of Figs. 2 and 3. For the wave packet with lowest energy, launched at $q_2 = 3.0$, these lines (labelled A to F in Fig. 2) turn out to be essentially those already observed in the spectrum of $P^e$, although with different intensities. Therefore, at this energy the low frequency oscillations of $|A(t)|^2$ primarily reflect the IVER dynamics of the model. Comparison of the two plots in Fig. 3 shows that this is no longer the case for the wave packet with highest energy launched at $q_2 = 0.5$. The lines associated with the oscillations of $P^e$ (labelled W to Z) are indeed quite weak in the Fourier transform of $|A(t)|^2$: in particular, they are weaker than another series of lines labelled p to u in the bottom plot of Fig. 3, none of these later lines having significant intensity in the power spectrum of $P^e(t)$. At this higher energy, the low frequency oscillations of $|A(t)|^2$ consequently do NOT reflect the IVER dynamics of the model. As indicated in Tables I and II, lines p to u are instead due to pairs of eigenstates which, to first approximation, can be described as $[0, v'_2, 0]$ and $[1, v'_2 - 2, 0]$, where $v'_2$ decreases from 17 (line p) to 12 (line u). Conclusion therefore is that at this energy the low frequency oscillations of $|A(t)|^2$ essentially reflect the detuning from 1:2 resonance between the frequency of the bend ($\omega'_2 + 2x'_{22}v'_2$) and that of the symmetric stretch ($\omega'_1$), the resonance being exact slightly above $v'_2 = 18$ for this model.

## IV - Relation to time resolved experiments

As already mentioned in the Introduction, the most intriguing result of time resolved spectroscopic experiments on $NO_2$ close to its first dissociation limit is indubitably the observation of a slow $NO^+$ transient signal with a particularly pronounced oscillatory component of period 600-800 fs, that is, an equivalent energy spacing of about 40-55 cm$^{-1}$ [13,14]. The origin of these oscillations is still a debated question, since it was suggested in Ref. [13] that these oscillations "may measure the average energy level spacing between the resonant levels of the $A^2B_2/X^2A_1$ states close to their conical intersection", while the authors of Ref. [14] tentatively assigned them to "wave packet motion along a very soft coordinate with an underlying harmonic level spacing of ~40 cm$^{-1}$", that is, "an essentially free rotation of an oxygen atom around a core NO molecule at large O-NO distances". The remainder of this paper is aimed at shedding some light on this question on the basis of the calculations presented above. More precisely, it is aimed at discussing the earlier proposition of Ref. [13]. Our model is indeed based on harmonic oscillators and consequently does not display the correct behaviour at large O-NO distances. Calculations based thereon are therefore not able to confirm nor to infirm the later hypothesis [14]. Still, we would like to mention that the authors of Ref. [14] invoke the rotation of an oxygen atom around the NO core at large O-NO distances because they consider that "this level spacing, 40 cm$^{-1}$, should be compared to the level spacing of the three normal vibrational modes of ground state $NO_2$ (1318, 750 and 1618 cm$^{-1}$ for the symmetric stretch, bend and asymmetric stretch, respectively)" [14]. However, the harmonic frequencies of $NO_2$ are no longer relevant at large O-NO distances. There, the meaningful quantities are instead the frequencies of the core NO stretch, the O-NO bend and the O-NO stretch. Since the frequency of the O-NO stretch precisely goes to zero at the dissociation threshold, this mode may be at the origin of the low frequency oscillations observed in experimental spectra. This hypothesis sounds more realistic than the rotation of the O atom around the NO core. Anyway, both tentative interpretations imply that a large portion of the initial excitation transfers very rapidly (*i.e.* within a few hundreds of fs) from the bend to the O-NO stretch degree of freedom, a scenario that still has to be ascertained (see below).

Except for the behaviour at large O-NO distances, the Hamiltonian of Eqs. (II-1)-(II-5) is expected to provide a good description of the two lowest electronic states of the $NO_2$ molecule. Moreover, the wave packet launched at $q_2 = 0.5$ is a correct approximation of the experimentally prepared wave packets, which consist of the almost vertical excitation of the vibrational ground state centred on $q_1 = q_2 = q_3 = 0$. Finally, the question of the nature of the experimental signal is admittedly a complex one, but it has been shown that in simple cases the observable ionization signal in pump-probe type experiments is directly related to the autocorrelation function $|A(t)|^2$ of the initial wave packet [23-25]. As a first approximation, we consider that this is also the case here. The fluctuations of $|A(t)|^2$ that were theoretically investigated in Sect. III therefore provide a good basis to discuss the hypothesis of the authors of Ref. [13], who suggested that the observed 40-55 cm$^{-1}$ angular frequency "may measure the average energy level spacing between the resonant levels of the $A^2B_2/X^2A_1$ states close to their conical intersection" [13].

A first comment is that wave packet calculations show that the angular frequency of the oscillations in $|A(t)|^2$ does not reflect the *average* energy level spacing of resonant levels but rather the energy gap between a *given pair* of eigenstates. It was also shown in Sect. III that the strongest lines in the Fourier transform of $|A(t)|^2$ form bands around the harmonics of the $\omega'_2$ fundamental frequency. This is the reason why experimentalists who investigate the time dynamics of diatomic molecules [26-28] as well as polyatomic ones [29-32] generally consider that the oscillations observed in the ionic or electronic signals reflect the frequency of a particular mode of the molecule. Here, things are however different because the period of the bending motion ($\approx$ 40 fs) is smaller than the temporal width of the laser ($\approx$ 90 fs [13,14]). All oscillations with frequency higher than 350 cm$^{-1}$, including all harmonics of $\omega'_2$, are thus strongly damped when the computed autocorrelation signal is smoothed over the laser width, as is the case for the experimental signal. This is clearly seen in Fig. 4, where the white traces show the time evolution of $|A(t)|^2$ smoothed with a gaussian window of width 90 fs (FWHM) and in Fig. 6, which displays a zoom of the smoothed signals in a narrower time window (0-2.5 ps). The principal components of the smoothed time evolution of $|A(t)|^2$ are thus the low frequency contributions that were discussed in detail in Sect. III. In particular, it was shown that, just above the conical intersection, the low frequency fluctuations of $|A(t)|^2$ essentially reflect the energy gap between pairs of states coupled by the non-adiabatic coupling, but that this is no longer the case at higher energies : for the wave packet launched at $q_2 = 0.5$, the leading contributions to low frequency fluctuations in $|A(t)|^2$ (lines p to u in the bottom plot of Fig. 3) indeed reflect the detuning from 1:2 resonance between the frequency of the bend ($\omega'_2 + 2x'_{22}v'_2$) and that of the symmetric stretch ($\omega'_1$). Of course, this result depends critically on the parameter $x'_{22}$ we introduced in the model (see Eq. (II-3)). For example, if one assumes that $x'_{22} = 0$ instead of $x'_{22} = -3$ cm$^{-1}$, then the detuning from 1:2 resonance is equal to $2\omega'_2 - \omega'_1 \approx 217$ cm$^{-1}$. Examination of Fig. 6 shows that, for $x'_{22} = 0$, the smoothed autocorrelation function accordingly exhibits (comparatively) high-frequency noise, while lines W to Z have insufficient intensity to generate noticeable low-frequency oscillations in the signal. At this point, it should be emphasized that recent experiments [33] confirm that $|x'_{22}|$ is of the order of a few cm$^{-1}$ and that it is consequently very plausible that the 1:2 resonance between the bend and the O-NO stretch becomes exact in the neighborhood of the dissociation threshold. Conclusion therefore is, that it is not very likely that the experimentally observed low-frequency oscillations in time domain spectra of NO$_2$ "may measure the average energy level spacing between the resonant levels of the A$^2$B$_2$/X$^2$A$_1$ states", nor even the energy gap between a given pair of resonantly coupled levels : the angular frequencies of the oscillations are more probably the energy gaps between states $[0, v'_2, 0]$ and $[1, v'_2 - 2, 0]$ that are strongly excited at time $t = 0$.

**V - Conclusion**

The wave packet calculations presented in this work confirm that there indeed exist two tentative explanations for the low-frequency oscillations observed in time domain spectra of $NO_2$ :

(i) they may reflect the existence of a low frequency mode, like the rotation of an oxygen atom at large distances from the NO core or, more probably, the O-NO stretch.

(ii) they may result from the fact that the symmetric stretch and the bend approach 1:2 resonance at the energy of the dissociation threshold.

Additional work is thus needed to determine which of these two phenomena is responsible for the signal with 600-800 fs period that is observed in time resolved spectra of $NO_2$ [13,14]. For example, the following points could be investigated :

- hypothesis (i) implies that a large portion of the excitation of the initial wave packet transfers rapidly (within a few hundreds of fs) from the bend to the O-NO stretch, while hypothesis (ii) requires in contrast that the energy remains essentially localized in the bend for at least several ps. It would therefore be of primary interest to check this point on surfaces that have the correct behaviour at large O-NO distances. Classical calculations presented in Ref. [19] suggest that energy indeed transfers very rapidly to the O-NO stretch degree of freedom, but the nonadiabatic coupling of the investigated model was 5 to 10 times too large and quantum mechanical calculations performed on the same model failed to reproduce this behaviour [19]. This is therefore a question which deserves further attention.

- one could try to find out whether the rotational motion invoked by the authors of Ref. [14] is plausible or not by looking at the periodic orbits on the lower electronic surface : the "rotational" motion of the wave packet is indeed possible only if there exists a not too unstable classical periodic orbit with the same characteristics.

- the evolution with increasing energy of the period of the oscillations could be investigated experimentally. Indeed, slow modes exist only close to the threshold and their period is an increasing function of energy. In contrast, detuning from the 1:2 resonance may lead to non-monotonous or even decreasing periods if the exact resonance is located below the dissociation threshold.

- last but not least, both the ionization process and the nature of the experimental signal are still debated questions [14]. No definitive conclusion regarding the origin of the slow periodic signal in time domain spectra of $NO_2$ will be met as long as these two points are not better understood. In particular, it was suggested in Ref. [14] that the slow $NO^+$ ions at the origin of the oscillations with 600-800 fs period may result from the absorption of two 266 nm photons by an almost dissociated molecule in its ground electronic state. If this were confirmed, then our assumption that the experimental signal is proportional to $|A(t)|^2$ would be invalidated, so that much more complex calculations would be needed to compare theoretical predictions and experimental results.

**Acknowledgments** : we are very grateful to Valérie Blanchet, Bertrand Girard and Béatrice Chatel (IRSAMC, Toulouse) for bringing the time resolved experimental results on $NO_2$ to our attention and for crucial discussions.

# TABLE CAPTION

**Table I :** Frequency (in cm$^{-1}$), period (in ps), and attribution of the most intense lines observed in the Fourier transforms of $P^e(t)$ and $|A(t)|^2$ in the range 0-120 cm$^{-1}$ (Figs. 2 and 3) and 0-2000 cm$^{-1}$ (Fig. 5). The two numbers in the last column indicate the respective ranks of the two eigenstates which are responsible for the existence of this peak in the Fourier transform (see Eqs. (III-5) and (III-7)). The energy and decomposition of all eigenstates that appear in this column are provided in Table II.

**Table II :** Energy (in cm$^{-1}$) and largest terms in the decomposition of the eigenstates which are responsible for the lines that appear in the Fourier transforms of $P^e(t)$ and $|A(t)|^2$, see Figs. 2, 3 and 5. The first column provides the rank of the eigenstate and the line(s) to which it contributes. In the last column, $(v_1, v_2, v_3)$ stands for the vector $|v_1, v_2, v_3\rangle$ of the harmonic oscillator basis of the ground electronic state and $[v'_1, v'_2, v'_3]$ for the vector $|v'_1, v'_2, v'_3\rangle$ of the harmonic oscillator basis of the excited electronic state.

**FIGURE CAPTIONS**

**Figure 1 :** Time evolution of the excited electronic state population, $P^e(t)$, for wave packets which at time $t=0$ are of minimum uncertainty and centred around $p_1 = p_2 = p_3 = q_1 = q_3 = 0$ and $q_2 = 3.0$ (dashed line) or $q_2 = 0.5$ (solid line) on the upper $A^2B_2$ electronic state. Potential energy at the centre of the wave packet at time $t=0$ is approximately 13800 cm$^{-1}$ for $q_2 = 3.0$ and 20800 cm$^{-1}$ for $q_2 = 0.5$.

**Figure 2 :** Squared modulus of the Fourier transform of $P^e(t)$ (top plot) and $|A(t)|^2$ (bottom plot) for the wave packet launched at $p_1 = p_2 = p_3 = q_1 = q_3 = 0$ and $q_2 = 3.0$ on the upper $A^2B_2$ electronic state. See Table I for the attribution of lines A to F.

**Figure 3 :** Squared modulus of the Fourier transform of $P^e(t)$ (top plot) and $|A(t)|^2$ (bottom plot) for the wave packet launched at $p_1 = p_2 = p_3 = q_1 = q_3 = 0$ and $q_2 = 0.5$ on the upper $A^2B_2$ electronic state. See Table I for the attribution of lines W to Z and p to u.

**Figure 4 :** Time evolution of the squared modulus of the autocorrelation function, $|A(t)|^2$, for wave packets which at time $t=0$ are of minimum uncertainty and centred around $p_1 = p_2 = p_3 = q_1 = q_3 = 0$ and $q_2 = 3.0$ (top plot) or $q_2 = 0.5$ (bottom plot) on the upper $A^2B_2$ electronic state. White traces show the same signals smoothed, however, with a gaussian window of width 90 fs (FWHM). Potential energy at the centre of the wave packet at time $t=0$ is approximately 13800 cm$^{-1}$ for $q_2 = 3.0$ and 20800 cm$^{-1}$ for $q_2 = 0.5$.

**Figure 5 :** Squared modulus of the Fourier transform of $|A(t)|^2$ for the wave packets launched at $p_1 = p_2 = p_3 = q_1 = q_3 = 0$ and $q_2 = 3.0$ (top plot) or $q_2 = 0.5$ (bottom plot) on the upper $A^2B_2$ electronic state. Zooms on the lowest frequency range (0-120 cm$^{-1}$) are shown in the bottom plots of Figs. 2 and 3.

**Figure 6 :** Time evolution of the squared modulus of the autocorrelation function, $|A(t)|^2$, smoothed with a gaussian window (90 fs FWHM) for wave packets which at time $t = 0$ are of minimum uncertainty and centred around $p_1 = p_2 = p_3 = q_1 = q_3 = 0$ and $q_2 = 3.0$ (dot-dashed line) or $q_2 = 0.5$ (solid line) on the upper $A^2B_2$ electronic state. The line with short dashes shows the time evolution of $|A(t)|^2$ for the wave packet launched at $q_2 = 0.5$ when the anharmonicity $x'_{22}$ is set to zero. Potential energies at the centre of the wave packets at time $t = 0$ are approximately 13800, 20800 and 21300 cm$^{-1}$, respectively.

**TABLE I**

| line | frequency (cm$^{-1}$) | period (ps) | eigenstates |
|---|---|---|---|
| **A** | 11.83 | 2.82 | 384 / 386 |
| **B** | 55.84 | 0.60 | 446 / 457 |
| **C** | 57.66 | 0.58 | 322 / 326 |
| **D** | 70.06 | 0.48 | 225 / 227 |
| **E** | 100.28 | 0.33 | 227 / 235 |
| **F** | 111.04 | 0.30 | 326 / 337 |
| **W** | 1.31 | 25.45 | 2058 / 2060 |
| **X** | 10.16 | 3.28 | 813 / 816 |
| **Y** | 16.09 | 2.07 | 1679 / 1683 |
| **Z** | 47.47 | 0.70 | 1857 / 1867 |
| **g** | 724.05 | 0.05 | 326 / 386 |
| **h** | 754.28 | 0.04 | 272 / 326 |
| **k** | 1478.33 | 0.02 | 272 / 386 |
| **p** | 22.94 | 1.45 | 1862 / 1867 |
| **q** | 27.78 | 1.20 | 1672 / 1679 |
| **r** | 49.19 | 0.68 | 1493 / 1503 |
| **s** | 53.63 | 0.62 | 1331 / 1345 |
| **t** | 67.71 | 0.49 | 1180 / 1193 |
| **u** | 72.36 | 0.46 | 1039 / 1059 |

**TABLE II**

| eigenstate | energy (cm$^{-1}$) | decomposition |
|---|---|---|
| 225 (D) | 11186.74 | 0.72(2,9,1)-0.58[0,0,2]+0.33[0,2,0] |
| 227 (D,E) | 11256.81 | 0.86[0,2,0]-0.34(2,9,1)+0.33(1,11,1) |
| 235 (E) | 11357.08 | 0.93(1,11,1)-0.31[0,2,0]-0.13[0,0,2] |
| 272 (h,k) | 11999.24 | 0.86[0,3,0]+0.33[0,1,2]+0.21(0,9,3) |
| 322 (C) | 12695.86 | 0.68(2,11,1)-0.46[0,4,0]+0.42[0,0,4] |
| 326 (C,F,g,h) | 12753.51 | 0.82[0,4,0]+0.46(2,11,1)-0.24(1,13,1) |
| 337 (F) | 12864.55 | 0.94(1,13,1)+0.23[0,4,0]+0.13[0,0,4] |
| 384 (A) | 13465.74 | 0.79(2,12,1)-0.42[0,1,4]-0.30[0,5,0] |
| 386 (A,g,k) | 13477.57 | 0.87[0,5,0]+0.41(2,12,1)-0.15(1,14,1) |
| 446 (B) | 14186.62 | 0.62(2,13,1)+0.55[0,6,0]+0.35[0,0,6] |
| 457 (B) | 14242.46 | 0.72(2,13,1)-0.58[0,6,0]-0.25[0,0,6] |
| 813 (X) | 17062.11 | 0.62[0,10,0]+0.50[1,0,8]+0.28(5,7,3) |
| 816 (X) | 17072.27 | 0.64[0,10,0]-0.48(5,7,3)+0.21[1,6,2] |
| 1039 (u) | 18395.21 | 0.77[1,10,0]+0.32(2,1,9)-0.25(0,22,1) |
| 1059 (u) | 18467.57 | 0.92[0,12,0]+0.21(1,16,3)-0.13[5,1,2] |
| 1180 (t) | 19086.48 | 0.57[1,11,0]-0.38(2,2,9)+0.29[2,1,8] |
| 1193 (t) | 19154.19 | 0.94[0,13,0]-0.19(4,16,1)-0.15(0,23,1) |
| 1331 (s) | 19783.70 | 0.87[1,12,0]+0.32(10,6,1)+0.16(1,22,1) |
| 1345 (s) | 19837.33 | 0.96[0,14,0]+0.11(1,22,1) |
| 1493 (r) | 20464.65 | 0.68[7,2,0]-0.46[1,13,0]+0.26(10,7,1) |
| 1503 (r) | 20513.84 | 0.92[0,15,0]+0.20(1,23,1)-0.12(0,25,1) |
| 1672 (q) | 21149.72 | 0.88[1,14,0]-0.19(8,7,3)+0.17(11,6,1) |
| 1679 (Y,q) | 21177.51 | 0.72[0,16,0]-0.42(1,24,1)+0.22(0,17,5) |
| 1683 (Y) | 21193.60 | 0.46(1,24,1)+0.40[2,10,2]+0.40[0,16,0] |
| 1857 (Z) | 21804.58 | 0.56(2,23,1)+0.55[8,2,0]+0.27[0,17,0] |
| 1862 (p) | 21829.12 | 0.87[1,15,0]+0.15[3,5,6]+0.14(4,15,3) |
| 1867 (Z,p) | 21852.05 | 0.89[0,17,0]-0.25(2,23,1)-0.17[3,9,2] |
| 2058 (W) | 22493.80 | 0.62[1,16,0]+0.45(12,6,1)+0.37[0,18,0] |
| 2060 (W) | 22495.11 | 0.68(12,6,1)-0.52[1,16,0]-0.32[8,3,0] |

**FIGURE 1**

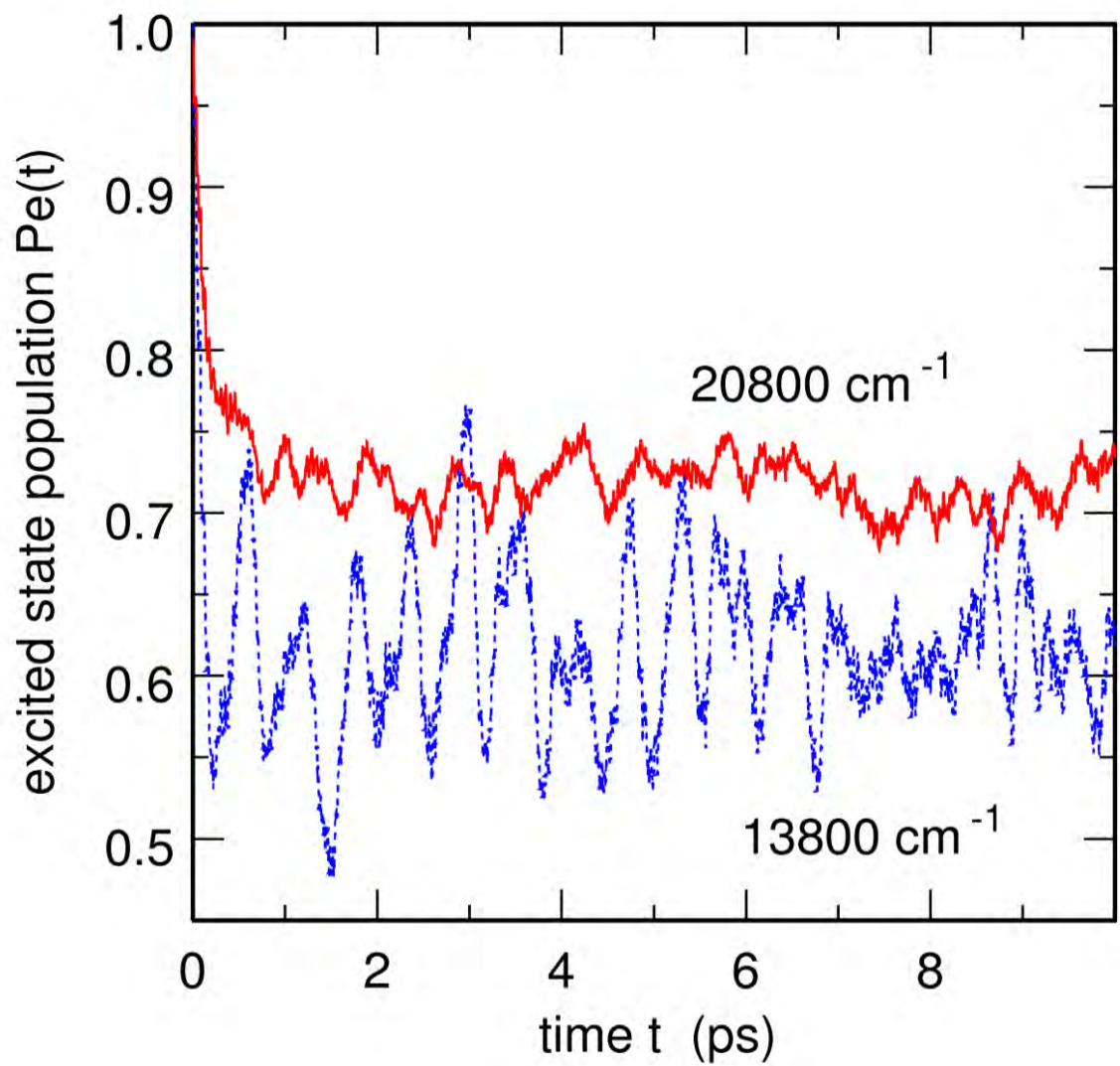

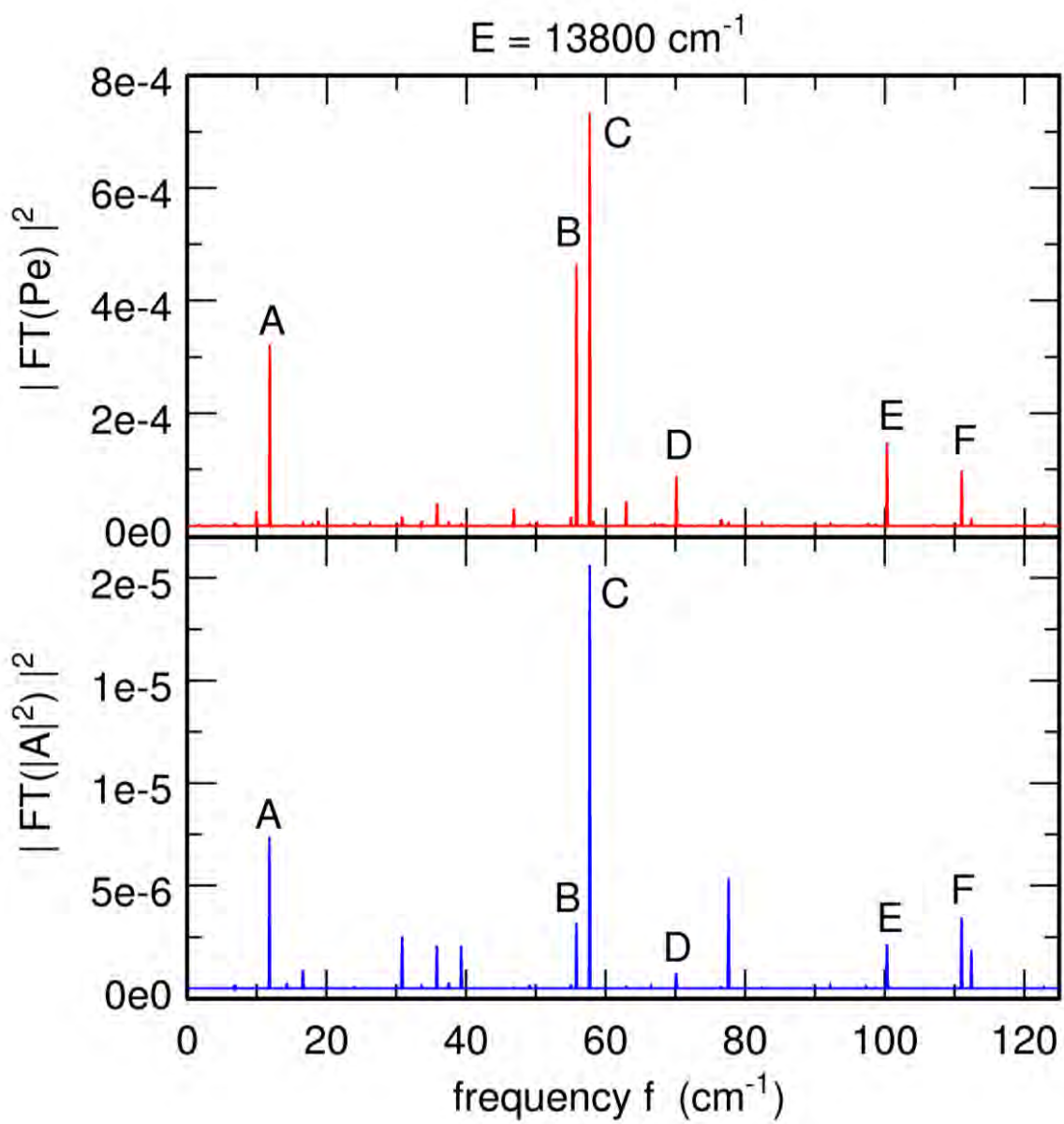

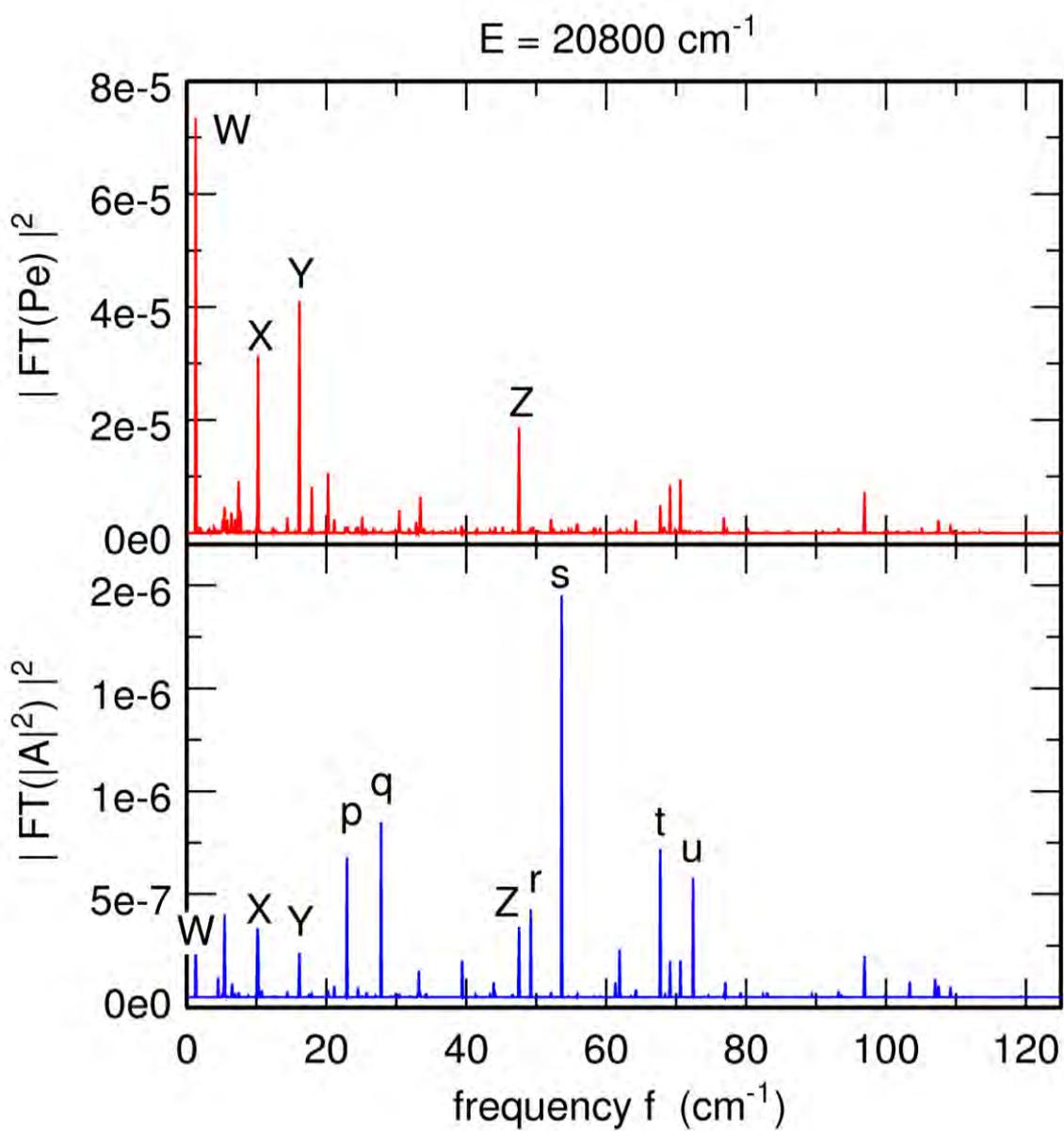

**FIGURE 4**

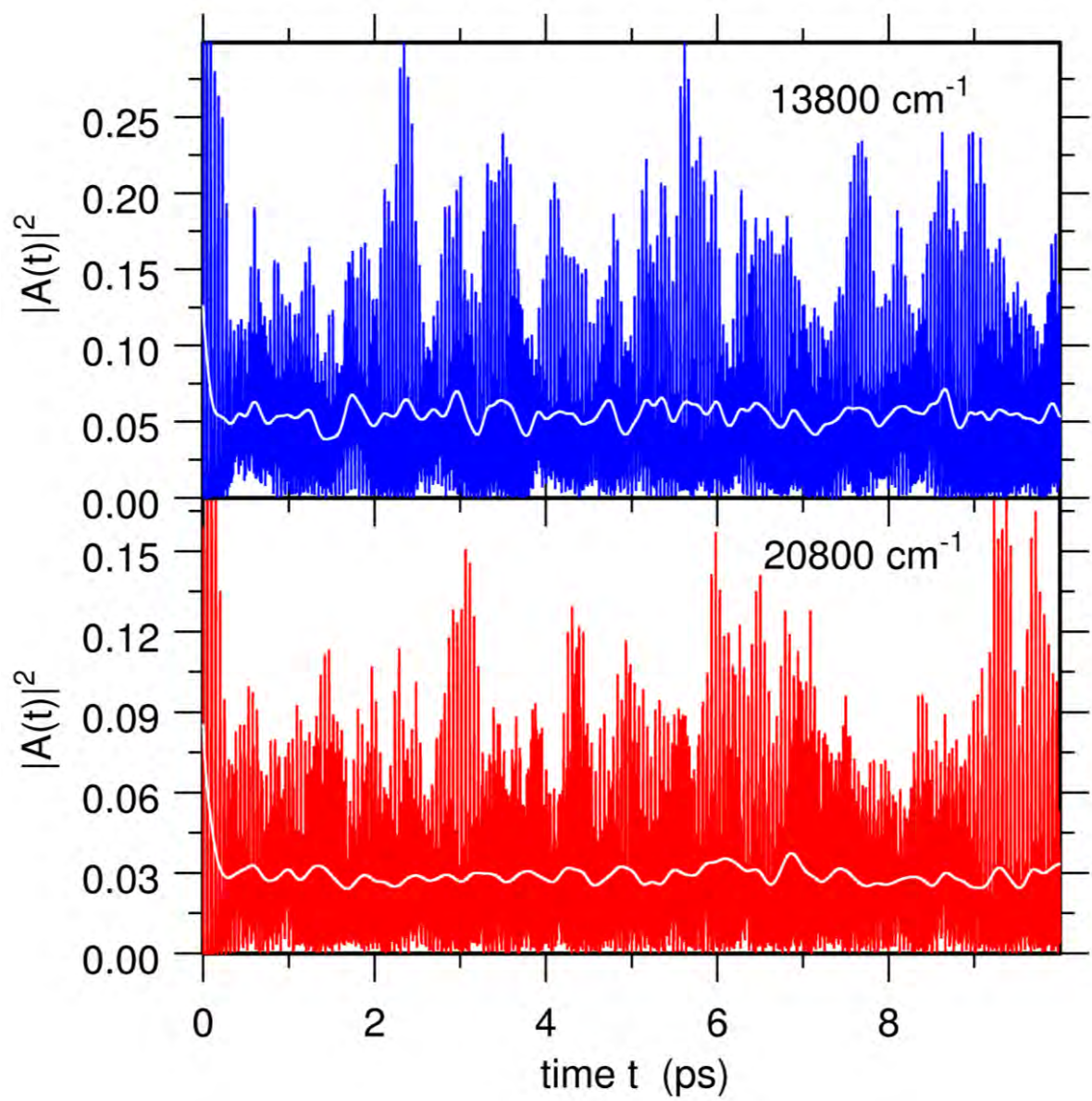

**FIGURE 5**

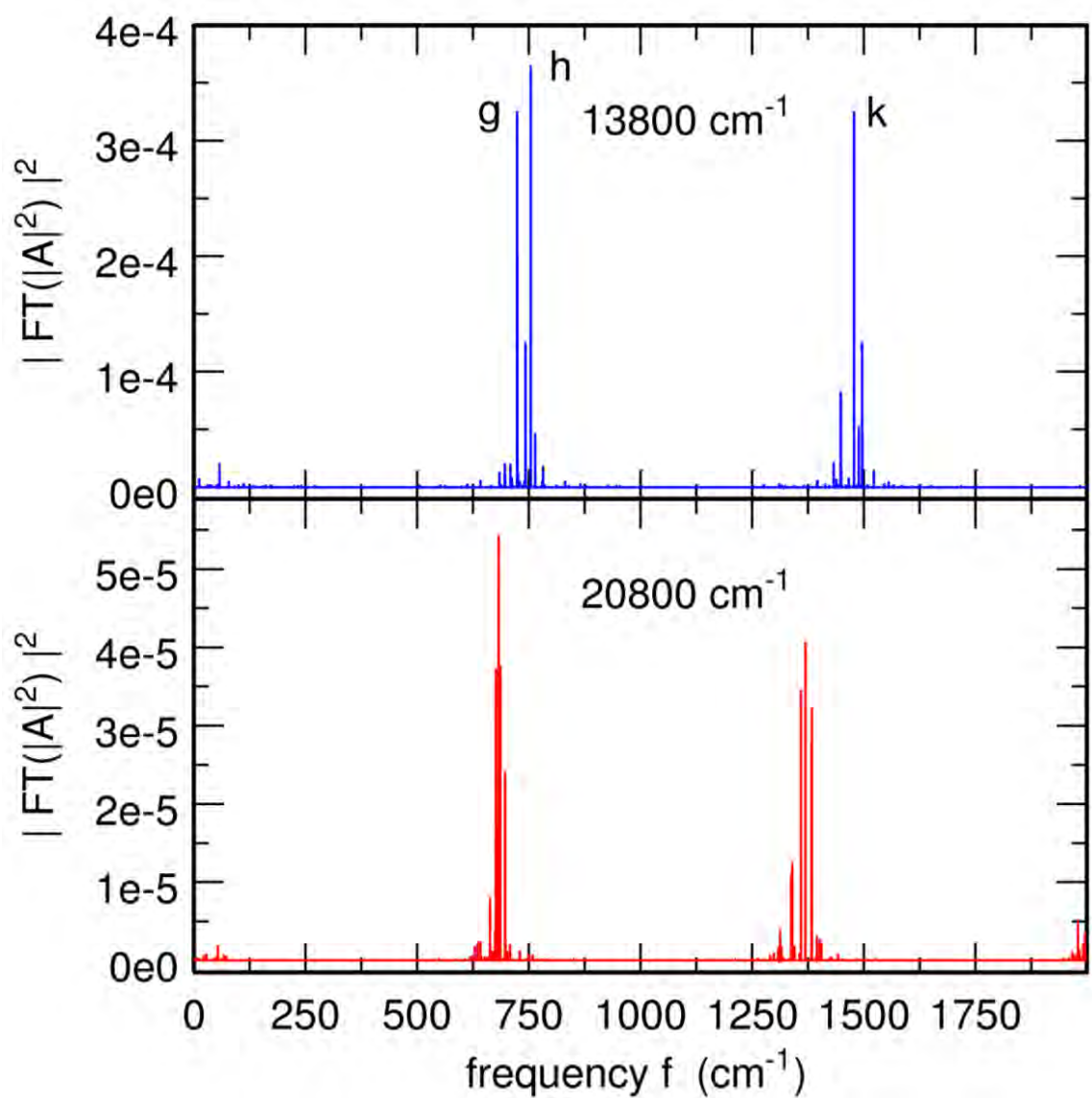

**FIGURE 6**

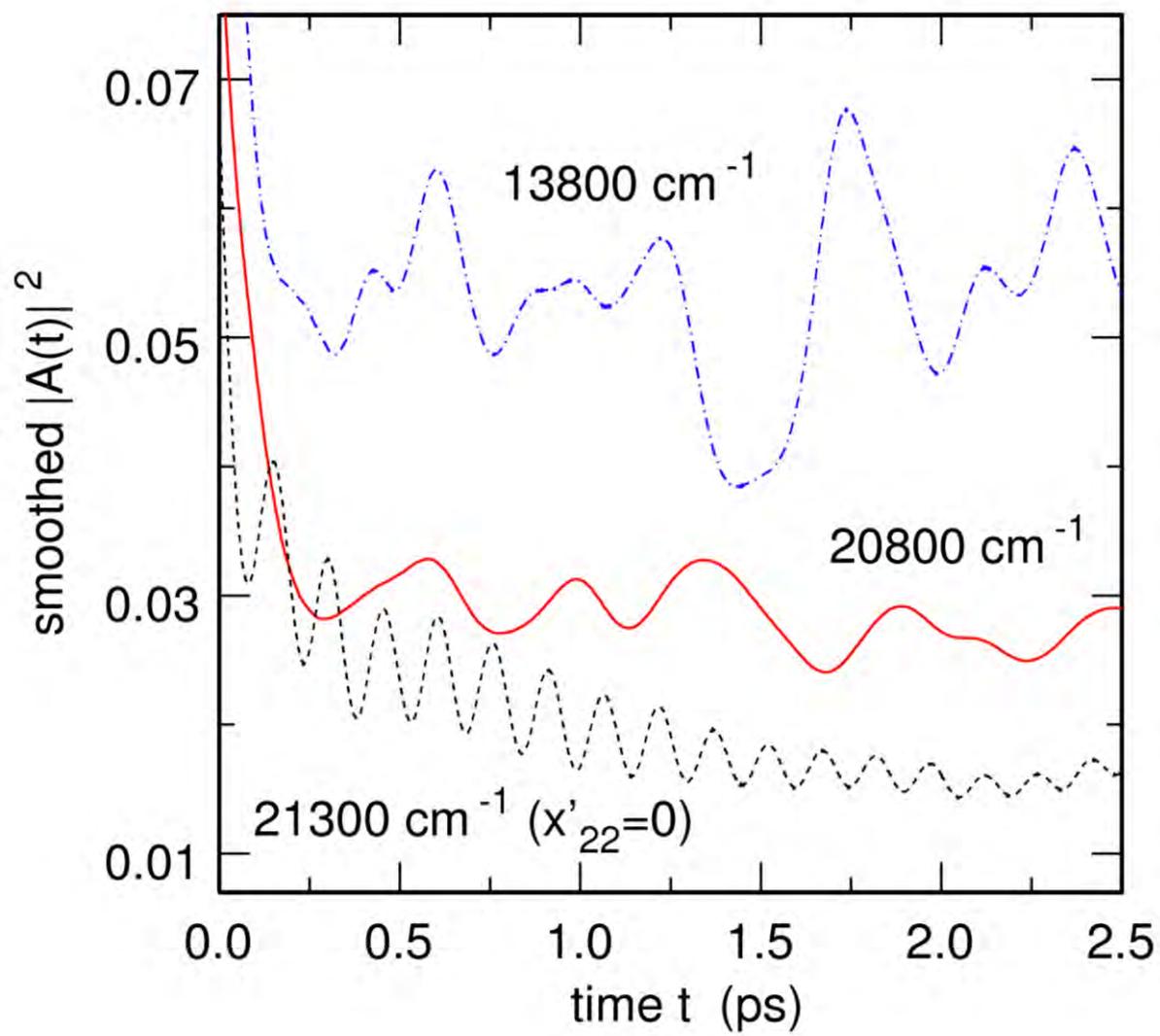